# Raman spectroscopy of active-carbon electrodes when Au colloids are placed at the electrolyte/electrode interface

H. Grebel[1] and Yuanwei Zhang[2]

[1] Center for Energy Efficiency, Resilience and Innovation (CEERI) and the ECE Department at the New Jersey Institute of Technology, Newark, NJ 07102. grebel@njit.edu

[2] Department of Chemistry and Environmental Science at the New Jersey Institute of Technology, Newark, NJ 07102. yuanwei.zhang@njit.edu

**Abstract:**

We use surface enhanced Raman spectroscopy (SERS) in studying functionalized Au nanoparticles (AuNPs) when incorporated in active-carbon (A-C) based super-capacitor cells. We observe a resonance-like enhancement in the graphitic line (G-line) vs the D-line (defect line) of the A-C electrode. We also observed an enhancement in the specific capacitance of super-capacitor cell as a function of AuNPs concentration. All of these may be explained by the formation of a quasi-2D array of AuNPs at the interface between electrolyte and the electrode.

# I. Introduction

Super-capacitors (S-C) are used in a wide-range of applications, such as consumer electronic products, memory back-up devices, hybrid electric vehicles, power supply system [1-2]. They were also proposed as buffers to highly fluctuating power grids that are equipped with sustainable sources [3]. They take advantage of the large capacitance at the narrow interface between a porous electrode and an electrolyte. Here we concentrate on carbon-based S-C that exhibit electrical double-layer [4-7]. Increasing the capacitance of super-capacitors may prove useful for these application.

To the S-C we added a low dispersion gold nanoparticles (AuNPs) – the mass ratio between the gold colloids and the active carbon is 1:5000 [8] – which resulted in a 10-fold enhancement in the gravitonic specific capacitance. The AuNPs are functionalized by a ligand that prevents coagulation while in suspension and ensures a strong electrostatic bond between the AuNPs and the electrode. That bond proved to be very strong upon washing the electrode. One ought to note that the Au colloids are not placed within the A-C particulate, but rather on its surface at a short distance from it. Placing the AuNPs inside the A-C particulate would result only in an impedance change of the electrode itself with little effect on the cell capacitance.

Generally, incorporating metal features into a dielectric material increases polarization of ordinary capacitors [9-13]. This was true low frequency regime of super-capacitors [8] and also true for the high-frequency regime - metal colloids exhibit Raman enhancement (Surface Enhanced Raman Spectroscopy, SERS [14-18]) and IR signals enhancement (Surface Enhanced IR Absorption, SEIRA, [19-20]) both demonstrated for an effective medium and for single elements. Periodic structures for SERS applications have been also investigated by us [21] and others, yet the combination of gold colloids and S-C is yet to be pursued. Ordinary Raman have been studied in electrochemical cells [22-23] and it would be natural to include SERS in the low and high frequencies.

Dispersions of the Au nano-particles (AuNPs) have been known for a long time. Their preparation methods are well-established, as well as, the relationship between their size, as determined by SEM or TEM and their optical scattering properties [24-25]. In order to ensure a good suspension, the colloids are functionalized with a negatively charged ligand. While the plasmonic peak absorption of uncoated AuNPs dramatically changes as a function of particles size [26], it is much less so for ligand coated particles [27]. A well-established technique that correlates the particle size to its optical scattering is dynamic light scattering (DLS) [28], which is used to determine the average particle size. The electrode itself is charge-neutral. Nevertheless, there is an electrostatic bond between the neutral, yet conductive A-C electrode and the negatively functionalized AuNPs. We use this bond to attach the AuNPs to the A-C electrode and use the ligand as a spacer between colloid and electrode.

For SERS under dry condition the larger the particle is better [29-30]. For super-capacitors, the capacitance effect takes place within the thin interface between the electrolyte and a solid electrode and one might wonder on the effect of the colloid's size. Using a full-wave analysis, the closer is the colloids to the electrode, the larger is the super-capacitive effect [31-32], similarly to SERS.

In this paper we propose that a quai-2D array, formed by the functionalized AuNPs at closed proximity to the A-C electrodes is responsible for the enhancement of the graphitic Raman line (G-line) over the D-line (defect line) in A-C electrodes, in addition to a 10-fold enhancement in the specific capacitance of the super-capacitor cells.

## II. Materials and Methods

### II.a Preparation methods - gold colloids:

A detailed description is provided in [8]. In short, in-house, AuNPs were synthesized by following a well-established method [33]; it employs citrate as a reducing agent and stabilizer. In brief, chloroauric acid ($HAuCl_4$) water solution (10 mg $HAuCl_4$ in 90 mL of water) was heated to boiling and sodium citrate solution (0.5 mL of 250 mM) was introduced. The mixture was stirred for 30 min until the color turned to wine red, or purple-brown. AuNPs were then purified by centrifuge and washed with DI water three times. The concentration of AuNPs in water was 1 mg/mL; the titration experiments were made with increasing amounts of 10 $\mu$L per batch and are translated to $\mu$g when referenced to the amount of A-C in the batch. The other batches were purchased from a vendor (Fisher) and their concentration was much smaller.

There were several batches of suspended AuNPs that have been used: commercial 40 and 100 nm (Thermo Fisher Chemicals, supplied in 0.1 mg/mL sodium citrate with stabilizer) and home-made (45 nm, [8], coated with sodium citrate). The 40 nm batch was concentrated via evaporation in order to reduce the ligand thickness and increase the Z-potential. The Zeta potentials for coated AuNPs were, -31.1 mV, -17.8 mV and -23.2 mV for the 45 nm, 40 nm (as is) and 100 nm (as is) and -33.6 mV for a concentrated batch of 40 nm colloids. This means that the Z-potential could be increased upon further concentration.

### II.b. Preparation methods - the porous electrodes:

A 100 mg of Cellulose Acetate Butyrate binder (CAB, Aldrich Chemicals) were first dissolved in 20 mL of acetone. 2 g of active-carbon (A-C, specific surface area of 1100 $m^2$/g, produced by General Carbon Company, GCC, Paterson, NJ, USA) were added and sonicated for 1 hour using a horn antenna. Vials, each containing 2 mL of the slurry were prepared. To these, succession amounts of 10 $\mu$L of AuNPs, suspended in water were added. Each mixture was further sonicated with the horn antenna for additional 30 min. The slurry was dropped casted on grafoil electrodes (area of contact 1.27x1.27 $cm^2$, manufactured by Miseal and purchased through Amazon), baked on a hot plate at <90 °C and then soaked with an electrolyte (1 M of $Na_2SO_4$, NaCl, or KOH). A fiberglass filter (Whatmen 1851-055), or a paper filter were used as a membranes. The Au colloid concentration were respectively, 1 mg/mL (home-made) and 0.1 mg/mL (Fisher). The Au colloids sizes were of order of ~40 and 100 nm (Fisher), 45 nm (home-made) as measured by dynamic light scattering (DLS). For the 40 and 100 nm colloids we used poly vinyl alcohol (PVA) as a binder (instead of the CAB). The PVA was dissolved in disulfide methyl sulfoxide (DMSO). Due to the lower concentration of the colloids, we added DMSO to the titration vials in order compensate for the added volume of colloid suspension in the sample.

On the average, the A-C particulates' dimension were ca 15 microns [31-32] and their specific surface area was rated as 1100 $m^2$/g.

### II.c. The samples:

Cuts of 200 micron thick grafoil electrodes with back adhesive (1.27 cm x 2.54 cm) were placed on similar cut microscope slides. Before placing it on the slides, the grafoil electrodes were heated for a few hours. The two slides were held by tweezers (or plastic clips) and the boundaries of the sample were left unsealed while soaking it in the electrolyte. The samples were later sealed with an epoxy. A hole was cut through the upper electrode for the Raman laser beam. The sample configuration is shown in Fig. 1a and its picture in Fig.1b.

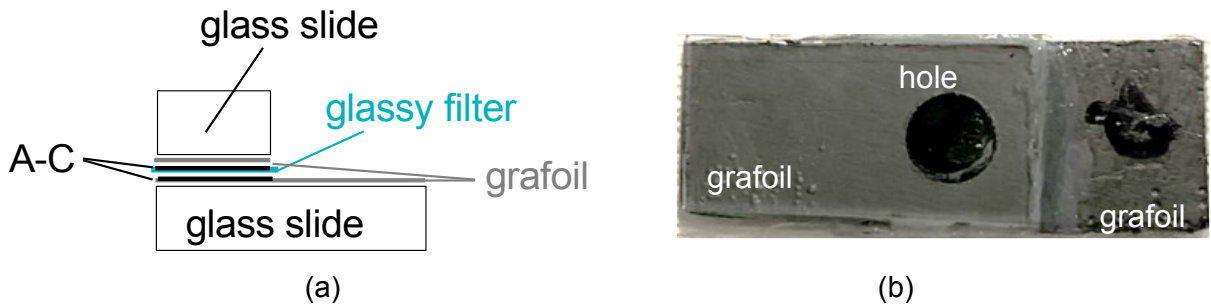

Figure 1.  (a) A cross section of sample with grafoil coated active-carbon (A-C) electrodes, and (b) a picture of the sealed sample, covering area of ca, 1.27x1.27 cm$^2$ with an entrance hole.  The number on the right indicates sample zero.

### II.d. Electrochemical Techniques:

Potentiostat/Galvanostat (Metrohm) was used in a 2-electrode set-up.  Most data here is provided with Cyclic Voltammetry (C-V) at scan rates of 100 mV/s.  As per [8], a large increase of the specific capacitance is observed for lower scan rates.  Charge-Discharge (C-D) at applied currents of 1 and Electrochemical Impedance Spectroscopy (EIS) between 50 kHz to 50 mHz completed the test.  The results all agreed on the titration trends.

### II.e. Raman Measurements:

The Raman system was made of 75-cm spectrometer equipped with a camera (cooled to -50 °C).  A Coherent laser, at $\lambda$=532 nm, with intensity of I=5 MW at the sample and x10 objective were used.  After moving averaging by 100 points, the data was fitted with two Lorentzian peaks.

### III. Results

### III.a. Raman spectra of the A-C electrode and the grafoil current collector:

In Fig. 2 we show a typical Raman spectra of the A-C and its current collector (grafoil).  The A-C exhibit two peaks in the range of 1200-1800: one at ca 1300 (the D-line) and the other at ca 1600 cm$^{-1}$ (the G-line).  The grafoil has a single G-line peak at 1575 cm$^{-1}$.

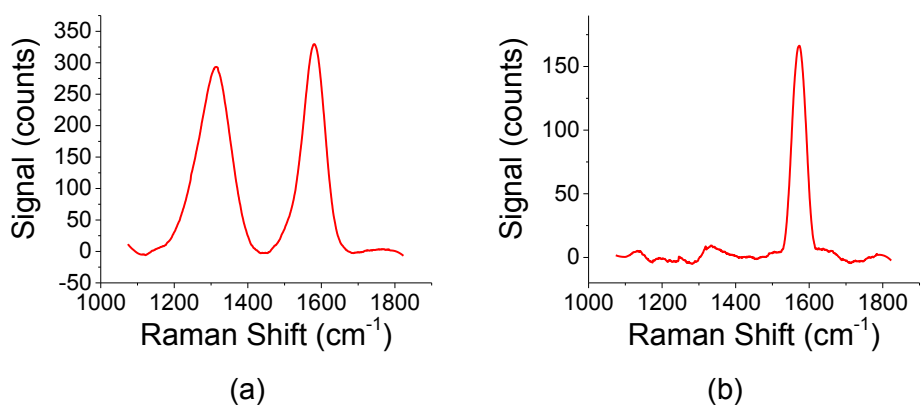

Figure 2.  Averaged Raman spectra of (a) dry A-C and (b) of the grafoil current collector

## III.b. As is 100 nm AuNPs:

Measurement on supercapacitors are made on bulk samples, thus averaging any local effect. In contrast, Raman measurement are local and could be affected by the number of local colloids and their separation (hot-spot). Figure 3a shows two Raman signals for AuNPs embedded 100 nm sample: one at some 'random' position, and the other for a 'preferred' position where the signal was maximized. Both signals were taken under dry conditions. The curves conveys two points: (a) the ratio $I_D/I_G$ is fairly constant regardless of the measurement conditions and (b) there is characteristic dip in the curve that accentuates the G-line at some AuNPs concentration. One ought to note that in the past, the ratio $I_D/I_G$ was used to determine the degree of defects in otherwise crystalline film - enhanced D-line peak over the G-line indicated a higher degree of an amorphous structure [34]

Under wet condition with an electrolyte, the peak at the peak at 70 μg of AuNPs per 200 mg of A-C in Fig. 3a has shifted to 50 μg of AuNPs per 200 mg of A-C. The sample has a PVA binder and 1 M $Na_2SO_4$ electrolyte. The specific capacitance did not show any trend within this range.

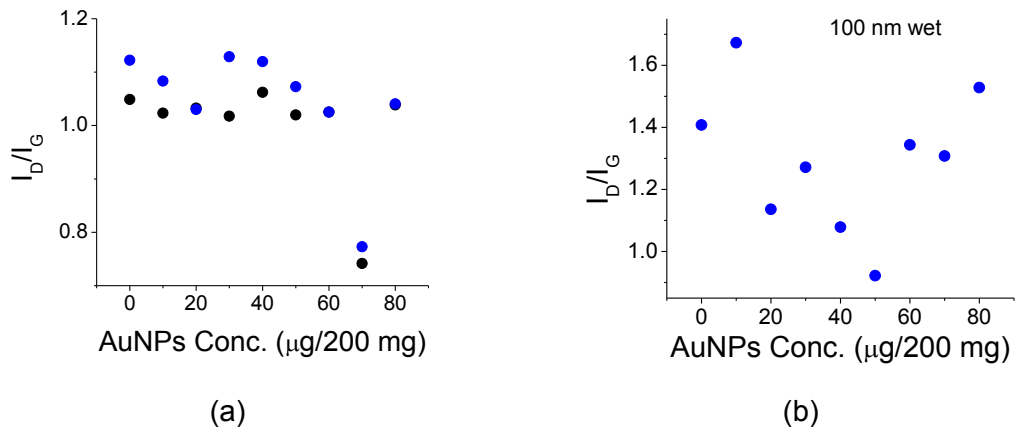

(a)    (b)

Figure 3.   A100 nm sample. (a) The ratio of $I_D/I_G$ for dry. Blue - preferred measurement position; Black - random measurement position. (b) $I_D/I_G$ ration for wet samples.

## III.c. As is 40 nm AuNPs:

40 nm AuNPs embedded cells are shown in Fig. 4. The peak of 70 μg of AuNPs per 200 mg of A-C shown in Fig. 3a has shifted to 30 μg/200 mg of A-C (Fig. 4a). The specific capacitance shows a clear transition at 30-40 μg/200 mg of A-C. The sample conditions were: C-V at 100 V/s; PVA binder; electrolyte – 1 M $Na_2SO_4$; glassy separator.

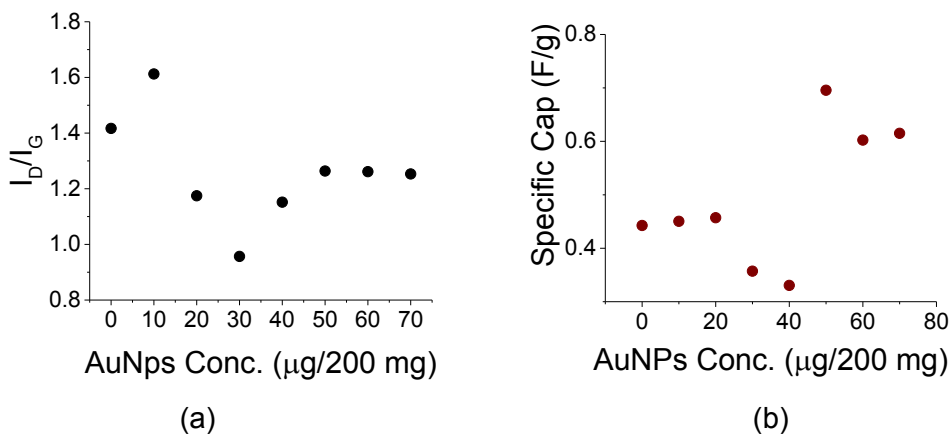

Figure 4. As is, un-concentrated 40 nm AuNPs sample. (a) The ratio of $I_D/I_G$ for a dry sample and (b) the gravitonic specific capacitance

### III.b. Concentrated 40 nm AuNPs:

Concentrated 40 nm colloids exhibited a larger Z-potential than the non-concentrated batch. They show a much more pronounced dip in the $I_D/I_G$ curves versus AuNPs concentration and a larger specific capacitance peak. The original batch was evaporated to reach a concentration of 0.5 mg/mL. Under wet conditions, Fig. 5a exhibits a minima in the $I_D/I_G$ curve at 10 µg/200 mg of A-C. The specific capacitance (Fig. 5b) exhibits a factor of 5 enhancement with respect to non-embedded AuNPs sample (zero concentration) with a clear maxima at 30 mg/200 mg of A-C. The sample conditions were: C-V at 100 V/s; PVA binder; electrolyte: 1 M NaCl; separator: paper filter).

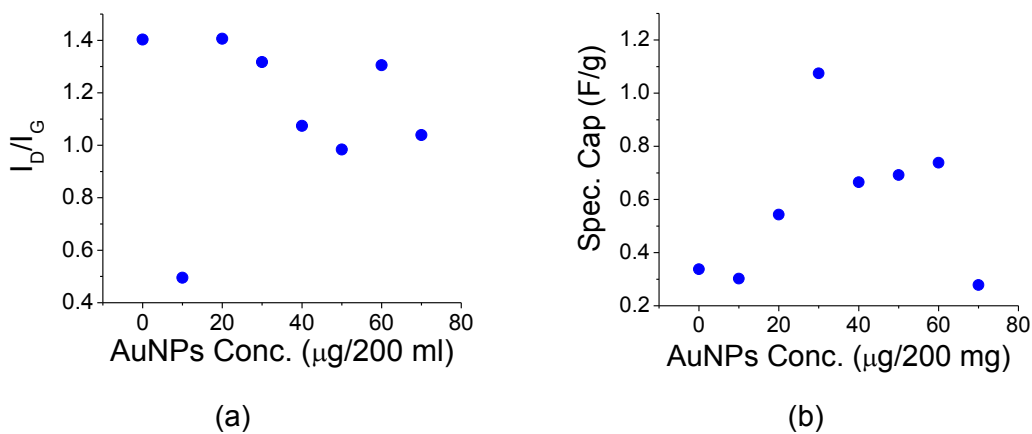

Figure 5. Concentrated 40 nm sample. (a) The ratio of $I_D/I_G$ under wet conditions. (b) The gravitonic specific capacitance.

### III.c. Home-made 45 nm AuNPs:

It seems that AuNPs with a larger Z-potential are better in terms of clarity of signals. Results for home-made, single-layer ligand samples are shown in Figs. 6-7. The concentration of the AuNPs was 1 mg/mL. As in [8]. The dry 45 nm sample exhibited a minima at $I_D/I_G$=30 µg/200 mg of A-C (Fig. 6a), whereas the wet sample exhibited a minima at a lower value, around $I_D/I_G$=10 µg/200 mg of A-C (Fig. 6b).  Condition: (C-V at 100 V/s; PVA binder; electrolyte – 1M $Na_2SO_4$; glassy separator).

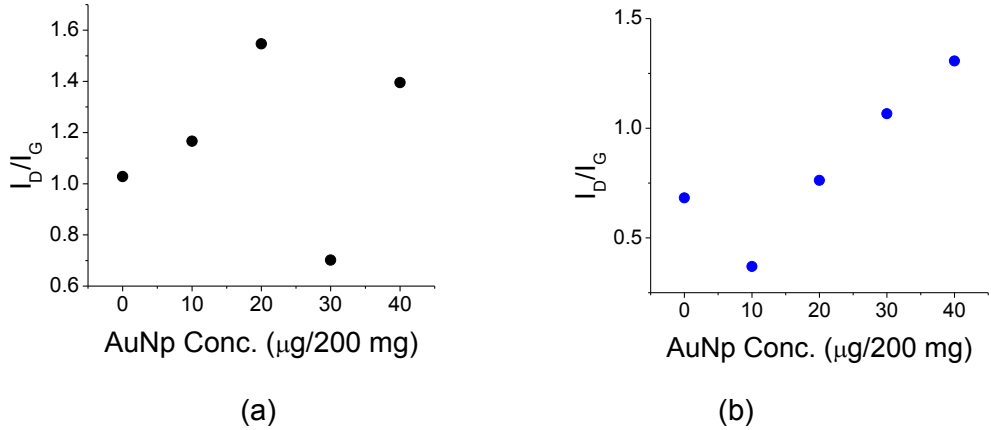

(a)          (b)

Figure 6. 45 nm sample.  (a) The ratio of $I_D/I_G$ under dry and (b) under wet conditions.

The gravimetric specific capacitance exhibited a peak at 30 µg/200 mg of A-C (Fig. 7a) and a peak for the G-line (Fig. 8b).  The, D-line exhibits a shift as a function of bias (Fig. 7c).

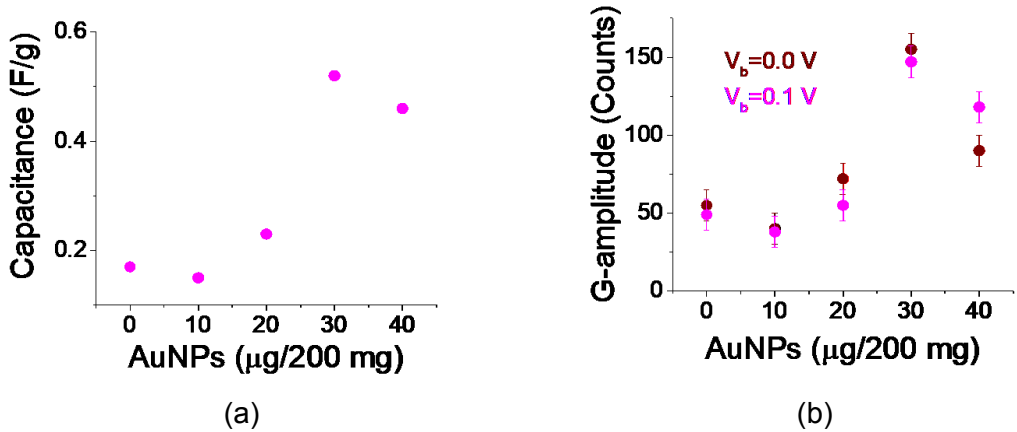

(a)          (b)

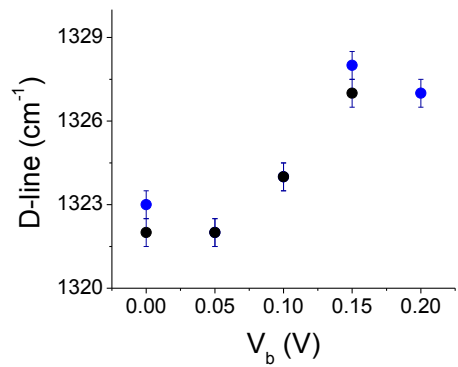

(c)

Figure 7. Home-made 45 nm sample. (a) Gravitonic specific capacitance as a function of titration. (b) The A-C G-line as a function of bias under wet conditions. (c) Change in the D-line of A-C as a function of bias: black curve - in the up direction; blue curve - in the down direction.

## IV. Discussion

We propose that a formation of a quasi-2D array by the functionalized AuNPs at closed proximity to the A-C electrodes explains the enhancement in the graphitic Raman line (G-line) over the D-line (defect line) of the A-C electrodes, in addition to the enhancement of specific capacitance in super-capacitor cells. The D-line of A-C is the result of interaction between the laser radiation and the charges that surrounds a defect point. The vibration is symmetric with respect to the defect point. The mode is absent from crystalline films. The G-line are the vibration of basal graphitic planes. The vibration is a-symmetric with respect to the planes and thus more susceptible to polarizations of the laser excitation [35-36]. Amorphous electrodes, such as active-carbon exhibit ratio of $I_D/I_G \sim 1$ (Fig. 3).

First we note the correlation between the Z-potential and the quality of resonance-like $I_D/I_G$ signal. Both concentrated 40 nm and home-made 45 nm batches exhibited a relatively larger Z-potential and a larger $I_D/I_G$ dip (meaning a larger enhancement of the G-line with respect to the D-line). At the same times, substantial enhancement of both D and G lines (Figs. 6b, 7b) resulted in a ratio of $I_D/I_G \sim 1$.

We also note that the AuNPs have similar enhancement effect whether the mass of the electrode is 100 mg/mL or 200 mg/mL. This means that the most, if not all of the AuNPs were bonded to the active carbon electrode and separated from it by the thickness of the functionalizing ligand layer. The ligand in suspension keeps the colloids apart, and we assume a quasi, close-pack array, as mimics by Fig. 8. Upon binding to the active-carbon electrode, the array maintains it quasi-periodic structure.

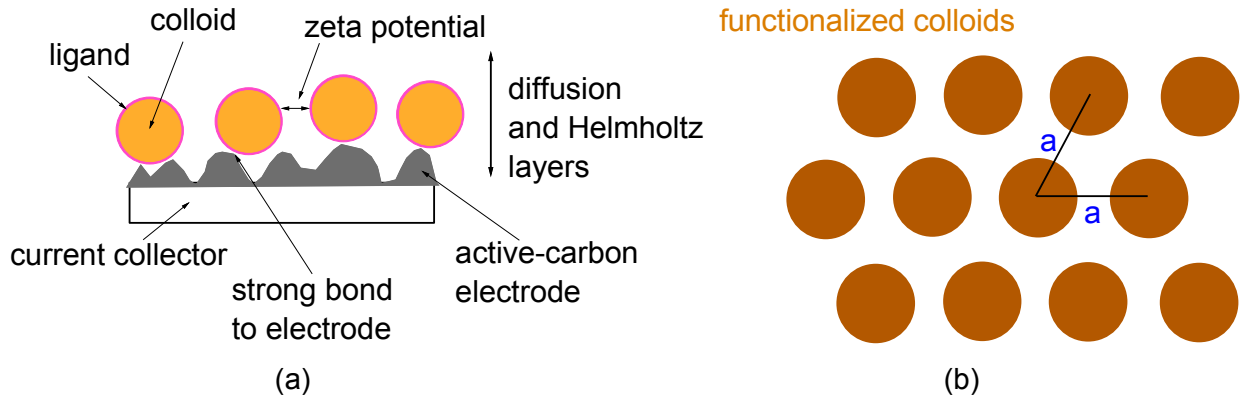

Figure 8. (a) Ligand coated colloids are forming a conductive array near the electrode (active carbon). (b) Quasi-2D array of AuNPs.

Selective signal-to-noise ratio (SNR) of scattered Raman lines may be associated with sub-wavelength resonances of a periodic structures. That resonance seems to favor the G-line over the other lines [37]. The wavenumber of the array depends on the electrolytic environment through its refractive index. If an optical beam is normally incident onto a surface mode and is also coupled to a standing mode via a Bragg scattering, then a resonance occurs.

$$(\lambda_0/a)\cdot[(4/3)\cdot(q_1^2-q_1\cdot q_2+q_2^2)]^{1/2}-n_{eff}=0. \quad (1)$$

Here, $\lambda_0$=532 nm - the laser excitation, "a" is the separation between colloids and $q_1$, $q_2$ are sub-integers related to the quasi periodic structure. The effective surface guide between colloids and the electrode is $n_{eff}$. With $n_{A-C}$, $n_{air}$ equal 1.88 and 1, respectively; the effective index of the electrode/air interface is, $n_{eff}=(n_{A-C}+1)/2=(1.88+1)/2=1.44$. When an electrolyte is present, $n_{electrolyte}$~1.3 and $n_{eff}=(1.88+1.3)/2=1.59$. Under sub-wavelength conditions there are many choices for $q_i$ so one picks the minimum periodicity that results in a resonance.

For the 1 mg/mL batch of Fig. 7, the average distance between colloids is computed for 30 μl as follows: (30 μL per batch)x(1 mg/mL)x(1/197 g/mol)x(6.02x10$^{23}$)~15x10$^{15}$ atom of gold; A 45 nm cluster contains ~2x10$^6$ gold atoms. Thus, the number of clusters is ~7x10$^9$ in a batch of 1 mL. If all colloids find themselves on the A-C electrode surface, then the average distance between them is ~1/√(7x10$^9$)~120 nm. At this distance, Eq 1 may be solved under dry conditions, $q_2$=0; $q_1$=1/4. Under wet conditions, the index change is satisfied by a smaller $q_i$ parameters, $q_2$=0; $q_1$=2, which translates to lower AuNPs concentration levels of 10 μL and a larger colloidal distance of 200 nm, as shown in Fig. 6a-b). From Figs 5 and 6, the relative and selective enhancement has a very narrow titration value alluding to scattering. At larger AuNPs concentrations (and smaller colloidal distances), local fields will amplify both the D- and G-Raman lines, as well as the cell capacitance. This is outlined below.

**Local field models:** We use a static model (Lorentz model) AuNPs-embedded S-C (Fig. 9). As the concentration of AuNPs increases, so is the effect of neighboring colloids on each other. Namely, the local field at some particular colloid position is a linear combination of the external field (the laser light, or the external cell bias) and an interaction field from all other neighboring colloids. The effect of all other colloids is characterized by an interaction parameter, C, which can be assessed by the potential of all other colloids on the colloid of interest (excluding the colloid itself). For a two dimensional case (one layer of colloid on the cell's electrode) the effect is proportional to the average distance between the colloids to the third power.

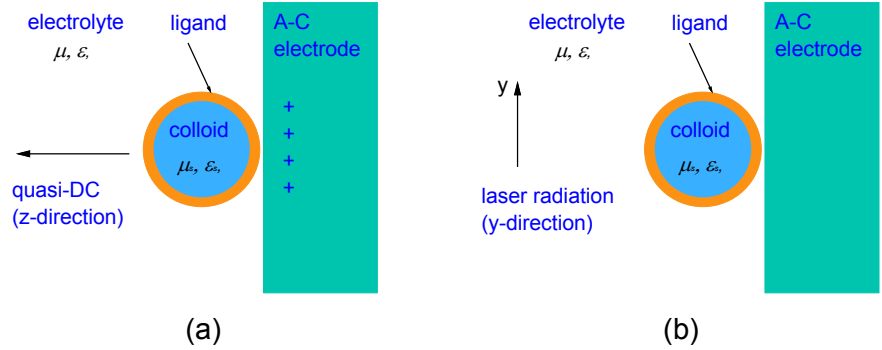

Figure 9. (a) Quasi-DC model. (b) SERS model.

**Low frequency model**: At very low frequencies, which are appropriate for S-C the electric field is perpendicular to the electrode(s) (Fig. 9a). The induced polarization on a spherical object is written as, $p=\varepsilon_0\alpha_e E_0$, with $\varepsilon_0$ - the vacuum dielectric constant; $\alpha_e$ - the colloid polarizability and $E_0$ is the external field between the anode and cathode electrodes. The local polarization is achieved in a self-consistent form, the Clausius-Mossotti relation and noting that the macroscopic polarization is P=Np. This is true when the distance between colloids is much smaller than the wavelength of interest and easily achievable at low frequencies of cell's operation [10]:

$$P=N\varepsilon_0\alpha_e E_0/(1-\alpha_e C). \qquad (2)$$

Along the z-direction, the coupling constant, $C_{LF}$, may be derived for two dimensional array of colloids, $C_{LF} \sim -2.4/\pi a^3$, with "a", the average distance between colloids. We ignore here a very small correction terms to $C_{LF}$. The polarizability of metals is $\alpha_e=4\pi r_0^3$, with $r_0$ being the colloid radius. When the colloid of a fixed radius $r_0$ fills the interface between the liquid and A-C electrode (including the diffusion layer), the density of colloids is, $N\sim(a^2 \cdot r_0)^{-1}$ The effective dielectric constant with respect to the electrolyte background (without the colloids) is $k=1+N\varepsilon_0\alpha_e E_0/(1-\alpha_e C)=1+4\pi r_0^3/r_0 a^2/(1+(2.4/\pi) \cdot 4\pi r_0^3/a^3)$. The effective dielectric constant is directly related to the capacitance increase in supercapacitors and is plotted in Fig. 10a. This model predicts an increase of ~2 in the cell capacitance, lower than is typically measured. The model is limited by $2r_0/a\sim1$. Beyond that point, the ligand prevents formation of a colloidal array and becomes erratic. The model predicts a 20% dielectric change and a larger concentration range when the AuNPs diameter is changed from 40 nm to 100 nm. Full wave simulations indicate that the large polarization spots are directly under the colloids - between the colloid and the A-C electrode [8].

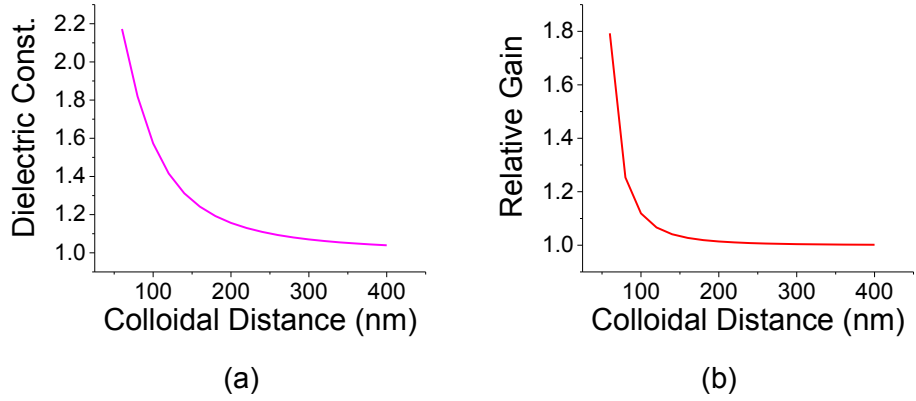

Figure 10. (a) Low-frequency model. The relative change in the dielectric constant as a function of colloidal distance in nm for a 45 nm colloids. The titration is typically conducted from large to smaller distances. (b) High-frequency model. Relative Raman gain coefficient as a function of colloidal distance in nm, for 45 nm colloids.

**High-frequency model**: The laser polarization is now along the y-direction, parallel to the electrode surface (s-polarization) (Fig. 9b). SERS has two enhancement factors: one from a single colloids and the other through the local field enhancement by all other colloids. The magnetic field is much smaller than the electric one at the near-field.

We follow the phenomenological approach for Raman scattering [30]: the *molecular* polarizability, $\alpha_m$ is expanded (one dimension) as, $\alpha_m = \alpha_{m0} + (\partial \alpha_m/\partial x)_0 (\delta x)$ to account for the (linear) molecular vibrations, $\delta x$. The interaction between the vibrating molecule (the A-C electrode) and the laser radiation, $E_0$, is written as: $p_{Raman} = \varepsilon_0 \cdot (\partial \alpha_m/\partial x)_0 \cdot (\delta x) \cdot E_0$. The vibration mode, $(\delta x)$, may be solved by use of induced harmonic oscillator, $(\delta x) = \varepsilon_0 \cdot (\partial \alpha_0/\partial x)_0 \cdot E_0 \cdot E_{Raman}$. In the presence of a single gold colloid, the NL polarization is enhance by a gain coefficient, $g \sim (r_0)^4$, with $r_0$ being the colloid's radius.

The effect of many conductive colloids on SERS may be added to each of the participating field component, incident and Raman components, through a polarization tern, $E_0^{local} = E_0 + p \cdot \alpha_e C_{HF}/\varepsilon_0$, and $E_{Raman}^{local} = E_{Raman} + p \cdot \alpha_e C_{HF}/\varepsilon_0$, respectively. Note the polarizability, $\alpha_e$, is of the colloids, not the molecules. The entire polarization is:

$$p = (p_{liear} + p_{SERS}) = \varepsilon_0 \alpha_e (E_0^{local} + E_{Raman}^{local}) + g \cdot (\varepsilon_0)^2 \cdot (\partial \alpha_m/\partial x)_0^2 \cdot (E_0^{local})^2 \cdot (E_{Raman}^{local}). \quad (3)$$

Frequency wise, while the NL Raman signal is shifted from the laser line, the intensity of each of the fields components, $E_0$ or $E_{Raman}$, is locally increased by the neighboring AuNPs, and therefore will affect the scattered signal - the Raman signal. In Eq. 3, we include the terms for the polarization, p, in a bootstrap fashion by using Eq.2 for single induced polarization. Concentrating on the last Raman term in Eq. 3, the Raman gain is modified by the local field,

$$g_R = N \cdot g \cdot (\varepsilon_0)^2 \cdot (\partial \alpha_m/\partial x)_0^2 \cdot (E_0)^2 \, [1 + \alpha_e C_{HF}/(1 - \alpha_e C_{HF})]^2. \quad (4)$$

For s-polarization of the excitation laser, the coupling constant $C_{HF} \sim 1.2/\pi a^3$ where a small correction term to $C_{HF}$ is ignored. The square brackets represent an enhancement factor per colloid per unit excitation intensity, and is plotted in Fig. 10b as a function of the distance between colloids. A substantial enhancement is predicted, yet, smaller than experimentally measured G-line (Fig. 7b). The model is valid up to a few titrations before failing (the point where the colloids touch each other, which also satisfies the condition, $1-\alpha_e C_{HF} \sim 0$). Beyond that point, the ligand prevents formation of a quasi-2D array and the signal decreases.

**V. Conclusion**

The inclusion of functionalized AuNPs in A-C based supercapacitors just above the electrodes, exhibited not only large amplification of specific capacitance, but also a large G-mode preference with respect to the D-line of Raman modes of active-carbon. The overall enhancement of the Raman A-C lines and the cell capacitance are explained through local field models, which depends on the colloidal dispersion. The selective enhancement of the electrodes' G-line over the D-line corroborates the formation of a quasi-2D array of AuNPs – the effect here is related to scattering by the periodic structure.